\documentclass[aip,apl,reprint,groupedaddress]{revtex4-1}
\usepackage{graphicx}
\usepackage{color}

\begin{document}
\title{An ultra-sensitive and wideband magnetometer based on a superconducting quantum interference device}
\author{Jan-Hendrik Storm}
\affiliation{Physikalisch-Technische Bundesanstalt (PTB), 10587 Berlin, Germany}
\author{Peter H\"ommen}
\affiliation{Physikalisch-Technische Bundesanstalt (PTB), 10587 Berlin, Germany}
\author{Dietmar Drung}
\affiliation{Physikalisch-Technische Bundesanstalt (PTB), 10587 Berlin, Germany}
\author{Rainer K\"orber}
\affiliation{Physikalisch-Technische Bundesanstalt (PTB), 10587 Berlin, Germany}

\date{\today}

\begin{abstract} 

The magnetic field noise in superconducting quantum interference devices (SQUIDs) used for biomagnetic research such as magnetoencephalography or ultra-low-field nuclear magnetic resonance is usually limited by instrumental dewar noise. We constructed a wideband, ultra-low noise system with a 45~mm diameter superconducting pick-up coil inductively coupled to a current sensor SQUID. Thermal noise in the liquid helium dewar is minimized by using aluminized polyester fabric as superinsulation and aluminum oxide strips as heat shields, respectively. With a magnetometer pick-up coil in the center of the Berlin~magnetically~shielded~room~2 (BMSR2) a noise level of around 150~aT~Hz$^{-1/2}$ is achieved in the white noise regime between about 20~kHz and the system bandwidth of about 2.5~MHz. At lower frequencies, the resolution is limited by magnetic field noise arising from the walls of the shielded room. Modeling the BMSR2 as a closed cube with continuous $\mu$-metal walls we can quantitatively reproduce its measured field noise.

\end{abstract}


\maketitle

Biomagnetism aims at the detection of magnetic fields generated by the human body. As these fields are typically in the range of femtotesla to picotesla when detected outside the human body, high sensitivity magnetometry is required. Traditionally, the preferred detectors are low critical temperature (low-$T_{c}$) superconducting quantum interference devices (SQUIDs) operated at liquid helium (LHe) temperatures. Owing to their exquisite sensitivity SQUIDs facilitated the measurements of magnetic fields of the brain, and commercial multichannel systems for magnetoencephalography (MEG) are available with a field noise of about 2~fT~Hz$^{-1/2}$. SQUID performance is usually limited by the LHe dewar due to Johnson noise in the superinsulation comprised of aluminized foils and the thermal radiation shields made from copper mesh.~\cite{Nenonen1996} Other noise sources, addressed in more detail below, include thermal noise of the human body, magnetic and thermal noise from the magnetically shielded room in which the system is usually operated, intrinsic SQUID flux noise and noise of the amplifier used for read out.~\cite{Koerber2016}
\par More recently, ultra-low-field magnetic resonance imaging (ULF MRI) has emerged which also utilizes SQUIDs as signal detectors.~\cite{Clarke2007,ULFNMR2014} In its most common variant, untuned superconducting pick-up coils are inductively coupled to current sensor SQUIDs. Until now, custom-designed SQUID systems with an improved noise have been built which utilize either low-noise commercial or home-built dewars, respectively. For instance, using second-order axial gradiometers, Zotev $et\,al.$~\cite{Zotev2007} achieve a minimal noise of 1.2~fT~Hz$^{-1/2}$ for a 37~mm diameter pick-up coil. Clarke $et\,al.$~\cite{Clarke2007} reach a total measured noise of 0.7~fT~Hz$^{-1/2}$ for a 63~mm diameter pick-up coil. 
Our single-channel system based on a 45~mm diameter first-order axial gradiometer has a total measured noise of 0.50~fT~Hz$^{-1/2}$ when operated inside a low noise dewar.~\cite{Fedele2015}

\begin{figure}[b]
\centering
\includegraphics[width=.95\columnwidth]{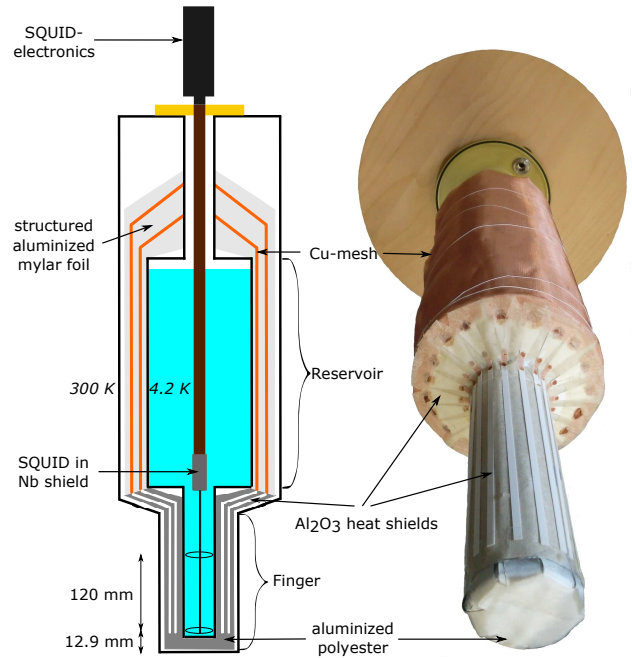}
\caption{\label{Fig:Figure1}Left: schematic setup of LINOD2 in gradiometer configuration. Right: view of one of the heat shields made from Al$_{2}$O$_{3}$ strips together with the copper mesh heat shield at the dewar reservoir. The outer shell has been removed.}
\end{figure}

\par It should be noted that atomic magnetometers have also been used for biomagnetic measurements albeit with a significant larger noise of about 10~fT~Hz$^{-1/2}$ and a bandwidth of about 100~Hz as demonstrated by Alem $et\,al.$~\cite{Alem2015} For a narrow bandwidth of about 10~Hz and a gradiometric setup inside a ferrite shield, Dang $et\,al.$~\cite{Dang2010} reached 160~aT~Hz$^{-1/2}$ at 40 Hz.
\par Improving the field sensitivity to the level of body noise, estimated to be around 50 aT~Hz$^{-1/2}$ for the torso \cite{Myers2007} would be of enormous benefit. For instance, in MEG high frequency components in evoked brain activity could be more easily resolved, paving the way for non-invasive studies of spiking activity.~\cite{Fedele2015,Storm2016} As we show below, improved sensitivity can be achieved by utilizing an ultra-low noise dewar and an increased pick-up coil. When expanding the approach to a multichannel system, larger coils place constraints on the localization accuracy of an activity at a depth $z$. To avoid aliasing in the spatial frequency domain, the center-to-center distance $D$ of the pick-up coils should fulfill $D\lesssim z$.~\cite{Ahonen1993a} For shallower sources one could potentially partly overlap the pick-up coils.
On the other hand, large pick-up coils are adequate for ULF MRI and allow an improvement of the currently poor signal-to-noise ratio. The enhanced image quality could potentially enable novel techniques such as neuronal current imaging~\cite{Koerber2016} and current density imaging.~\cite{Vesanen2014}

\par In this work we report on further improvements of our ultra-low noise SQUID system operated inside a Low Instrinsic NOise Dewar (LINOD). The dewar has a volume of 6.5 liters and is based on the design of Seton $et\,al.$ who used aluminized polyester as superinsulation and aluminium oxide as the heat shield material. This proved to be successful for a tuned system operated at 414~kHz.\cite{Seton2005} In our first approach LINOD1, aluminized polyester was also partly used as superinsulation but the heat shields were entirely made from copper mesh resulting in an equivalent field noise of 0.45~fT~Hz$^{-1/2}$ of the dewar for a 45~mm diameter first-order axial gradiometric pick-up coil.~\cite{Fedele2015} The upgrade to LINOD2 involves changes to the finger section containing the pick-up coil as shown in Fig.~\ref{Fig:Figure1}. In the finger and the cone section adjacent to the LHe reservoir, the superinsulation consists of aluminized polyester and the heat shields are made from commercially available strips and plates of aluminum oxide (LCP GmbH, Hermsdorf, Germany). The separate plates were connected via strips of copper mesh which were glued with GE Varnish to ensure good thermal contact. In the remaining part we used structured aluminized Mylar foil and copper mesh as before. When the dewar is cold, the minimal distance of the pick-up coil to the outside of the bottom was measured as 12.9~mm. With the SQUID system installed, the hold-time of the dewar is about 4 days. The average boil-off rate in the first 3.5 days is 1.45 liters per day measured with a vibrating membrane dip-stick.

     \begin{table}[t]
 \caption{\label{Tab:Table1}SQUID parameters and white noise levels for the two setups with 45~mm diameter pick-up coils. The magnetic flux is given in units of the flux quantum $\Phi_{0}$.}
 \begin{ruledtabular}
 \begin{tabular}{l c c}
	Parameter	& Gradiometer & Magnetometer \\\hline
$I_{\rm{c}}$\,$(\mu \rm{A})$$^{a}$ & 7.3 & 6.7 \\
$R_{\rm{N}}$\,$(\Omega)^{a}$ & 11 & 9.8 \\
$C$\,$(\rm{pF})$$^{a,b}$ & $ 0.4$ & $ 0.4$ \\
$L_{\rm{SQ}}$\,(pH)$^{b}$ & $ 80$ & $ 80$ \\
$L_{\rm{i}}$\,(nH) & 400 & 400 \\
$M_{\rm{i}}$\,(nH) & 3.97 & 4.0 \\
$L_{\rm{p}}+L_{\rm{str}}$\,(nH) & 413 & 270 \\
$B_{\Phi}$\,(pT/${\Phi_{0}}$) & 265 &  220 \\
$S_{\Phi,\rm{i}}^{1/2}$\,($\mu\Phi_{0}$~Hz$^{-1/2})^{c}$ & 0.52 & 0.485 \\
$V_{\Phi}$\,($\rm{m}V/{\Phi}_{0}$) & 1.07 & 0.763 \\
$S_{B,\rm{amp}}^{1/2}$\,(aT~Hz$^{-1/2})^{d}$ & 92 & 107 \\
$S_{B,\rm{i}}^{1/2}$\,(aT~Hz$^{-1/2}$) & 138 & 107 \\
$S_{B,\rm{m}}^{1/2}$\,(aT~Hz$^{-1/2}$) & 166 & 151 \\
$\epsilon_{c,\rm{i}}$\,($h)^{e}$ & 22 & 19 \\
$\epsilon_{c,\rm{m}}$\,($h)^{f}$ & 32 & 38 
 \end{tabular}
 \end{ruledtabular}
\begin{flushleft}
$^{a}$ referred to the Josephson junction \\
$^{b}$ estimated from the layout \\
$^{c}$ intrinsic SQUID noise with SQUID inductance screening \\
$^{d}$ includes amplifier and wiring contributions \\
$^{e}$ referred to total intrinsic noise $S_{B,\rm{i}}^{1/2}$ \\
$^{f}$ referred to total measured noise $S_{B,\rm{m}}^{1/2}$
\end{flushleft}
 \end{table}

\par  The single-channel system utilizes a current sensor SQUID inductively coupled to a superconducting pick-up coil of inductance $L_{\rm{p}}$ forming an untuned coupling scheme. The double transformer layout described in detail in Drung $et\,al.$~\cite{Drung2007} was used.
The critical current $I_{\rm{c}}$, the normal state resistance $R_{\rm{N}}$, and the capacitance $C$ of the Josephson junctions are listed in Tab.~\ref{Tab:Table1} together with the SQUID inductance $L_{\rm{SQ}}$ and the input coil inductance $L_{\rm{i}}$. 

\par For the evaluation of the system, two individual sensor probes equipped with different 45~mm diameter pick-up coils were used: A single-turn, first-order axial gradiometer with a baseline of 120 mm and a single-turn magnetometer, respectively. Each was connected to the input coil of a separate single-stage current sensor SQUID. For this setups the coupled energy sensitivity per unit bandwidth $\epsilon_{c}=S_{\Phi}L_{\rm{i}}/(2M_{\rm{i}}^2)$ is used as the figure of merit which takes into account both the intrinsic flux noise $S_{\Phi}^{1/2}$ and the mutual inductance $M_{\rm{i}}$ between the input coil and the SQUID. The equivalent magnetic flux density noise $S_{B}^{1/2}$ can be calculated by $S_{B}^{1/2}=S_{\Phi}^{1/2}L_{\rm{tot}}/(M_{\rm{i}}A_{\rm{p}})=(2\epsilon_{c}/L_{\rm{i}})^{1/2}L_{\rm{tot}}/A_{\rm{p}}$ where $A_{\rm{p}}$ is the pick-up loop area and $L_{\rm{tot}}=L_{\rm{p}}+L_{\rm{str}}+L_{\rm{i}}$ is the total inductance of the input circuit with $L_{\rm{str}}$ representing stray inductances of the interconnection lines. $L_{\rm{tot}}$ was determined as described by Storm $et\,al.$~\cite{Storm2016} Increasing the pick-up coil diameter $d$ improves $S_{B}^{1/2}$. For a single turn pick-up coil and $L_{\rm{i}}=L_{\rm{p}}$ one finds $S_{B}^{1/2}\propto d^{-3/2}$.

\par The other noise contribution to be taken into account stems from the room-temperature amplifier used to read out the SQUID. In terms of flux noise it is given by $S_{\Phi,\rm{amp}}^{1/2}=S_{V,\rm{amp}}^{1/2}/V_{\Phi}$ where $V_{\Phi}$ is the flux-to-voltage transfer coefficient at the SQUID working point and $S_{V,\rm{amp}}^{1/2} = 370$\,pV~Hz$^{-1/2}$ is the measured voltage noise from the amplifier including wiring. In Tab.~\ref{Tab:Table1} the main parameters of the two setups are given.

\par
\begin{figure}[t]
\centering
\includegraphics[width=\columnwidth]{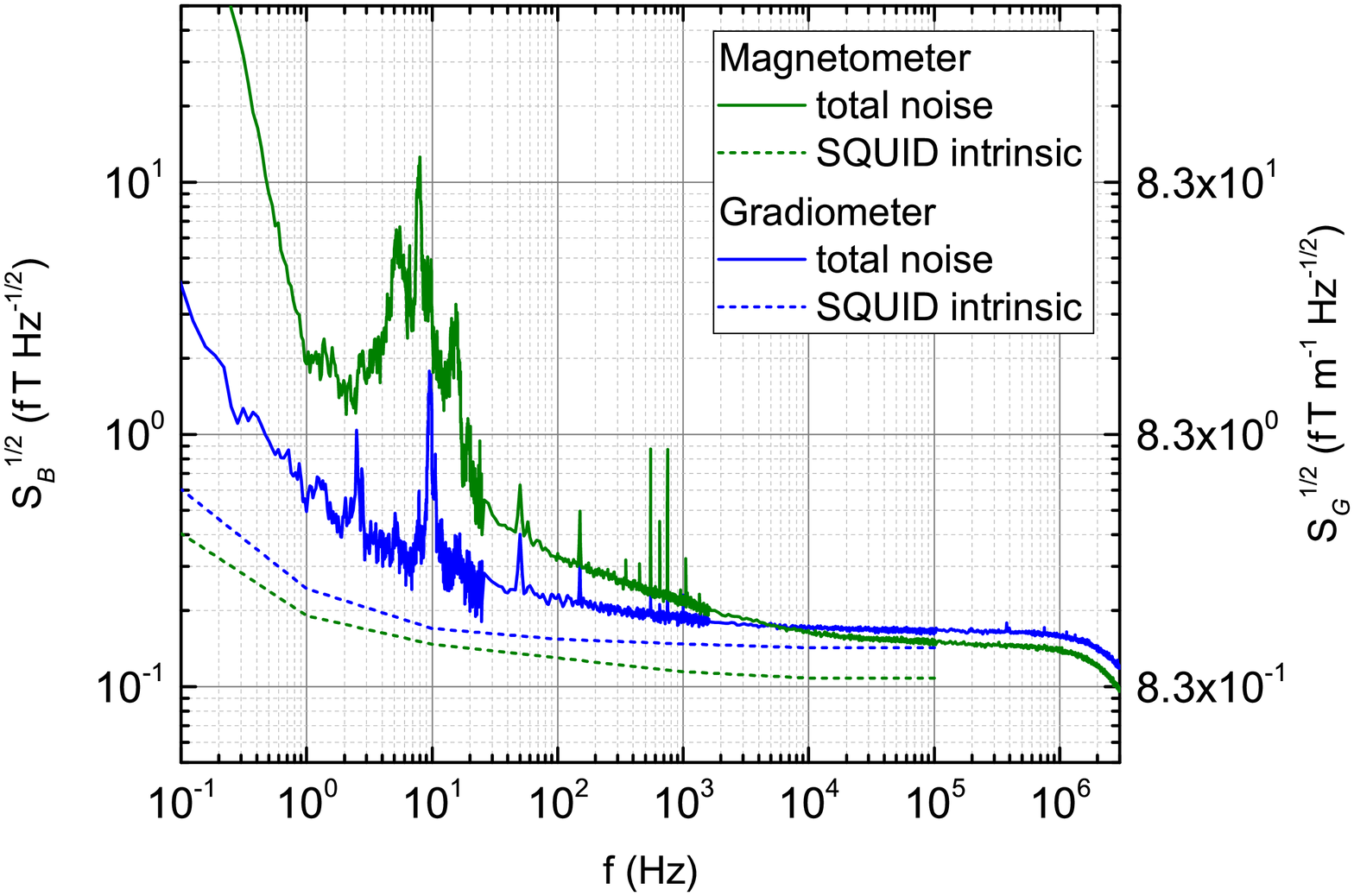}
\caption{\label{Fig:Figure2}Measured magnetic flux density noise $S_{B,\rm{m}}^{1/2}$ for the two setups with 45~mm diameter pick-up coils. Magnetometer (solid green curve), gradiometer (solid blue curve). The calculated intrinsic SQUID noise levels $S_{B,\rm{i}}^{1/2}$ are given by the dotted curves. For the gradiometer, the noise is referred to the bottom pick-up loop, and the gradient noise is shown on the right.} 
\end{figure}

 For characterization we measured the field noise of the single-channel SQUID systems when operated in the center of the Berlin magnetically shielded Room 2 (BMSR2).~\cite{Bork2000} In Fig.~\ref{Fig:Figure2}, the total measured noise $S_{B,\rm{m}}^{1/2}$ for the gradiometer and the magnetometer and also the gradient noise $S_G^{1/2}$ are shown. The gradiometer and magnetometer show a white noise of about 170~aT~Hz$^{-1/2}$ and 150~aT~Hz$^{-1/2}$, respectively. The 3-dB bandwidth is about 2.5 MHz for both setups. The pronounced peaks in the gradiometer spectra at 2.5~Hz and 10~Hz are probably due to resonances in the dewar mounting used for the experiment. This is supported by separate measurements in another shielded room with a different mounting scheme for which these resonances were not observed. 
\par The intrinsic field noise $S_{B,\rm{i}}^{1/2}$ of the SQUIDs for the gradiometer and magnetometer amounts to about 140~aT~Hz$^{-1/2}$ and 110~aT~Hz$^{-1/2}$, respectively (dotted lines in Fig.~\ref{Fig:Figure2}). $S_{B,\rm{i}}^{1/2}$ corresponds to the minimally achievable noise and was determined from the SQUID flux noise, measured without the pick-up coils connected. The inductance screening~\cite{Clarke2004} of $L_{\rm{SQ}}$ by the pick-up coil inductance was considered for the calculation. The corresponding intrinsic coupled energy sensitivities $\epsilon_{c,\rm{i}}$ amount to 22\,$h$ and 19\,$h$ for the gradiometer and magnetometer, respectively.
These values are close to the theoretical limit~\cite{Clarke2004} $\epsilon_c=16 k_{\rm{B}} T (L_{\rm{SQ}} C)^{1/2} /k^2=16\, h$ obtained at the temperature $T=4.2$ K. Here, $k_{\rm{B}}$ is the Boltzmann constant and $k\approx 0.7$ gives the coupling coefficient between the input coil and the SQUID.

\par
\begin{figure}[t]
\centering
\includegraphics[width=\columnwidth]{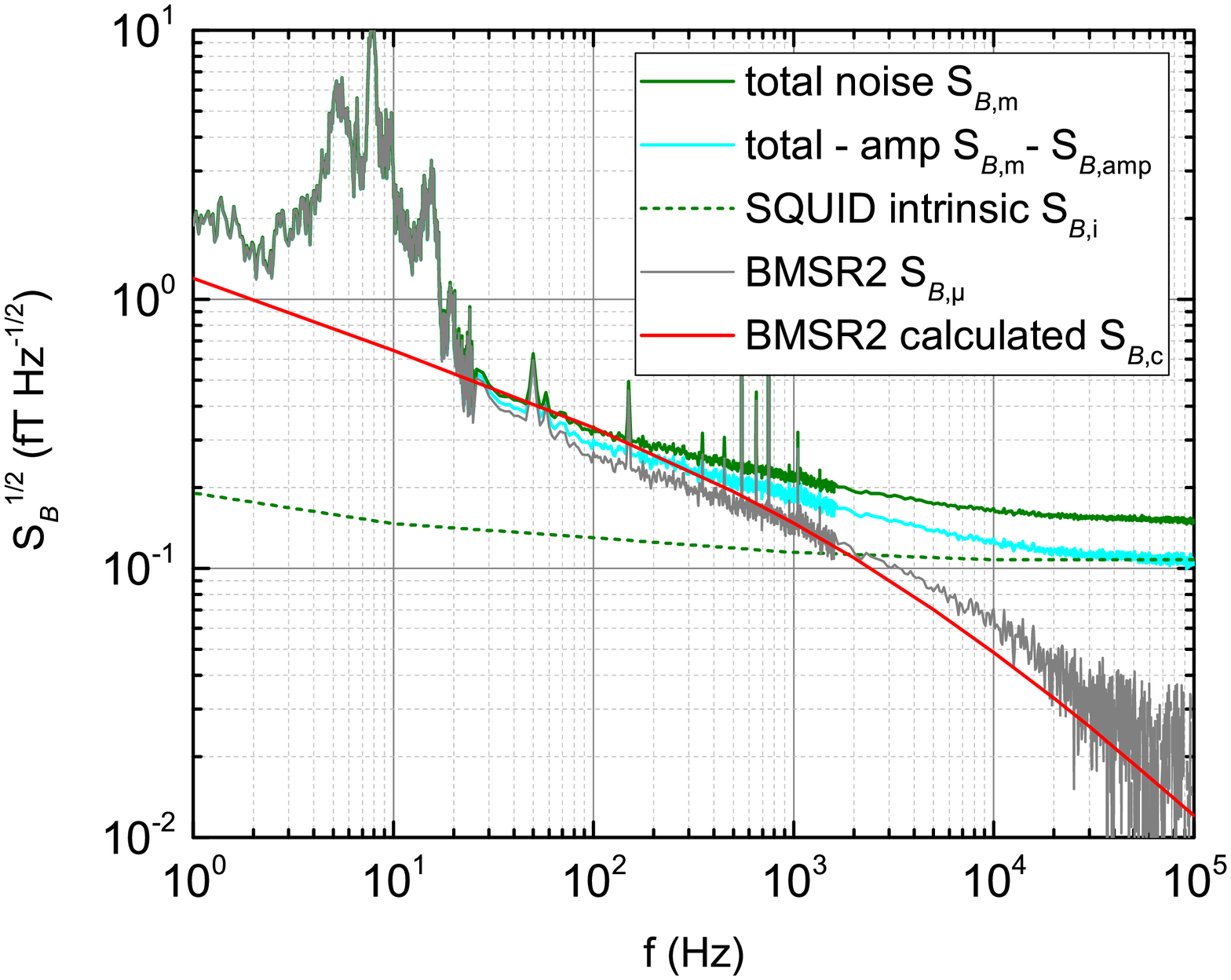}
\caption{\label{Fig:Figure3} Breakdown of the noise contributions of the magnetometer together with a comparison of the measured magnetic noise of the $\mu$-metal walls $S_{B,\mu}^{1/2}$ and the results from the numeric calculation $S_{B,\rm{c}}^{1/2}$.}
\end{figure}

 By comparing the magnetometer and the gradiometer noise, it is obvious that there is an additional far-field noise source seen only by the magnetometer. In Fig.~\ref{Fig:Figure3} all individual noise components of the magnetometer are shown. If $S_{B,\rm{amp}}$ is subtracted from $S_{B,\rm{m}}$ we reach the intrinsic SQUID contribution $S_{B,\rm{i}}$ at about 50~kHz. This shows that SQUID and amplifier noise are the sole contributions above this frequency. Another noise component $S_{B,\mu}$ is extracted via $S_{B,\mu}=S_{B,\rm{m}}-S_{B,\rm{amp}}-S_{B,\rm{i}}$. It originates from the innermost $\mu$-metal walls of the BMSR2 and can be quantitatively measured between $\approx 30$\,Hz and $\approx 50$\,kHz. At 100~Hz in the center of the room a noise level of about 260~aT~Hz$^{-1/2}$ is determined. We also find $S_{B,\mu}^{1/2}\propto f^{1/4}$ from approximately 100~Hz up to about 1~kHz and a roll of above. At frequencies below about 30~Hz, mechanical vibrations and background field fluctuations dominate the measurement.

\par Calculations for Johnson and thermal magnetic field noise of conducting and ferromagnetic materials were performed by several authors.~\cite{Lee2008,Nenonen1996} In order to explain $S_{B,\mu}(f)$ measured with the magnetometer, we computed the field noise of the shielding environment according to the fluctuation-dissipation theorem.~\cite{Callen1951} In this way, $S_{B,\mu}(f)$ can be calculated from the dissipated power in the shielding walls generated by a time harmonic current $I_0 e^{i \omega t}$ in the pick-up loop of the magnetometer. From the time averaged power $P(f)=I_{0}^{2} R_{\rm{eff}}(f)/2$ an effective resistor can be determined which links the dissipated power in the walls to the square of the voltage fluctuations per unit bandwidth  $S_{V}(f)=4k_{\rm{B}}TR_{\rm{eff}}(f)$. By making use of the principle of reciprocity~\cite{Hoult1976} and Faraday's law, the flux density noise of the shielding environment is given by:
\begin{equation}
S_{B}^{1/2}(f)= \frac{\sqrt{4k_{\rm{B}}T \cdot 2P}} {2 \pi f A_{\rm{p}} I_{0}}.
\end{equation}
To obtain the dissipated power for our experimental setup, we used FEM methods to calculate the spatial electrical and magnetic field distributions. The eddy-current losses $P_{e}$ are given by the volume integral of the electrical field $\vec{E}$ over the shielding walls:
\begin{equation}
P_{e} = \frac{1}{2}  \int_v \sigma\, \vert\vec{E}\vert^{2} dv. \label{eq.2}
\end{equation}
The power dissipation due to magnetic losses $P_{m}$ is given by the volume integral of the magnetic field $\vec{H}$:
\begin{equation}
P_{m} = \pi f  \int_v \mu^{''} \vert \vec{H} \vert^2 dv. \label{ep.3}
\end{equation}
Here $\sigma$ is the electrical conductivity and $\mu ''$ is the imaginary part of the permeability $\mu=\mu ' +i\mu ''$ of the $\mu$-metal walls.

\par The geometry used for the calculation was a closed cube with an inner edge length of 3.2~m and a wall thickness of 4~mm. This corresponds to the innermost $\mu$-metal layer of the BMSR2 and gives a good approximation for $f>1$~Hz due to the skin-effect. The pick-up coil of the magnetometer was placed in the center of the cube, the surface normal pointing along the vertical direction of the room. Exploiting the symmetry of the geometry, only 1/8 of the whole geometry was needed in the FEM model which reduces the calculation time. We used typical values for the $\mu$-metal: $T=293\,\textrm{K}$, $\sigma=1.5\times 10^6\,\Omega^{-1} \textrm{m}^{-1}$ and $\mu/\mu_0 =  45000+i1800$. This corresponds to a loss tangent of  $\mu ''/\mu ' =0.04$.
We calculated the total magnetic noise $S_{B,\rm{c}}$ of the BMSR2 as the sum of the eddy current and the magnetic contributions (red line in Fig.~\ref{Fig:Figure3}). Above 1~Hz, $S_{B,\rm{c}}$ is essentially given by the eddy current contribution which is about a factor of five larger than the magnetic counterpart.
In the frequency range from 20~Hz to about 20~kHz, we find a good agreement between the measured and calculated data. The slight disagreement can be the result of the simplified geometry and a somewhat overestimated conductivity in the model. For instance, mechanical stress after assembly of the $\mu$-metal walls could reduce the values for $\sigma$ and $\mu$.

In summary, the design of Seton $et\,al.$~\cite{Seton2005} for ultra-low noise LHe dewars utilizing aluminized polyester as superinsulation and aluminium-oxide as heat shields is also suitable for untuned SQUID systems leading to negligible dewar noise contributions. With a magnetometer configuration we achieve an extremely low white noise of about 150~aT~Hz$^{-1/2}$ which contains a significant contribution from the readout amplifier. Below 20~kHz we are limited by magnetic noise emanating from the innermost $\mu$-metal walls of the BMSR2. This is not seen in the gradiometer setup yielding a better sensitivity in the range below 3~kHz. With FEM simulations treating the inner walls as a continuous and closed cube we obtain good agreement with the measured field noise using the magnetometer setup. To further improve the performance, the amplifier noise contributions can be minimized by employing a two-stage readout scheme enabling a white field noise of about 140~aT~Hz$^{-1/2}$ and 110~aT~Hz$^{-1/2}$ for the gradiometer and magnetometer, respectively. Beyond that, a reduction of the SQUID noise is necessary which could potentially be achieved by utilizing sub-micrometer Josephson junctions.~\cite{Schmelz2017} Apart from the superior noise performance it should be noted that the high bandwidth makes this system well suited for many applications also outside biomagnetism. We intend to use this system, amongst other, for ULF MRI and in particular for the realization of current density imaging and neuronal current imaging.

\begin{acknowledgments}
The authors thank Allard Schnabel and Eva Al-Dabbagh for contributions during construction of LINOD2.\\
This work has received funding from the European Union's Horizon 2020 research and innovation programme under grant agreement No 686865 and by the DFG under grant No KO 5321/1-1.
\end{acknowledgments}

\bibliography{bibliography}

\end{document}